\documentclass[11pt,twoside]{article}
\usepackage{hurlburt}
\usepackage{epsf,graphicx,amsmath}
\begin{document}
\title{Effects Of Granulation Upon Larger-Scale Convection}
\vspace{-0.2truein}
\author{N. E. Hurlburt\altaffilmark{1}, M. L. DeRosa\altaffilmark{1}, K. C. Augustson\altaffilmark{2}, and J. Toomre\altaffilmark{2}}
\altaffiltext{1}{Lockheed Martin Solar and Astrophysics Laboratory, 3251 Hanover Street, Palo Alto, CA 94304}
\altaffiltext{2}{JILA and Dept. of Astrophysical \& Planetary Sciences, University of Colorado, Boulder, CO 80309}

\begin{abstract} 
We examine the role of small-scale granulation in helping to drive supergranulation and even larger scales of convection. The granulation
is modeled as localized cooling events introduced at the upper boundary of a 3-D simulation of compressible convection in a rotating 
spherical shell segment. With a sufficient number of stochastic cooling events compared to uniform cooling, we find that supergranular
scales are realized, along with a differential rotation that becomes increasingly solar-like.
\vspace{-0.2truein}
\end{abstract}

\section{Introduction}
The solar differential rotation profile exhibits prominent radial shear layers near the top and bottom of the convection zone \citep{thomp96}.
The near-surface shear layer (NSSL) occupies the upper 5\% of the Sun by radius, whereas the tachocline begins near the base of the convection 
zone. Much of the dynamics of the NSSL is due to the vigorous granular-scale convection that is driven by radiative cooling and large 
superadiabatic gradients. The collective interaction of these granular-scale flows (average sizes of $1$~Mm, lifetimes of $0.2$~hr) is a major
component in the formation of supergranular ($15$-$35$~Mm, $24$~hr) and mesogranular ($5$-$10$~Mm, $5$~hr) scales \citep[e.g.][]{rast03,nord09}.
The preferential influence of the Sun's rotation upon supergranular and larger scales may lead to the anisotropic angular momentum transfer seen
in the NSSL. The overturning times and sizes of supergranules suggest that granular convection is weakly influenced by rotational effects. The 
interactions between the small-scale convection, larger velocity gradients, and rotation yield Reynolds stresses which are significant in 
transporting angular momentum within the layer \citep{derosa02}. 

A direct numerical simulation of all these dynamical scales is not possible given current computational resources. Therefore, in order to 
investigate the dynamics of the $100$-$1000$s of supergranules necessary for adequate statistics, we must model the small scales that we cannot 
capture. We examine two methods of carrying the required energy flux through the top of the domain, one using a spherically symmetric (uniform) 
cooling function while the other is a granular-scale stochastic cooling model. In this paper, we investigate the dynamics and differential rotation 
profiles achieved in our simulations. The governing equations and numerical approach used in solving them are briefly discussed in \S \ref{sect2}.
Our granulation model is reviewed in \S \ref{sect3}, while the connection between these simulations and the near-surface shear layer of the solar 
convection zone is discussed in \S \ref{sect4}.

\section{Formulating the Problem\label{sect2}}
As a compromise between physical accuracy and more rapid time evolution, a perfect gas and a radiative diffusion approximation are employed in 
our simulations. The spherical segment domains used in our simulations involve large portions of the Sun's inherent spherical geometry, which 
is necessary to properly capture the effects of rotation on supergranular scales. To model these dynamics below the solar photosphere, we use a 
heavily modified version of a fully compressible 3-D hydrodynamic simulation code. The Curved Shell Segment (CSS) code is a mature modeling tool
which solves the Navier-Stokes equations of motion in rotating spherical segments \citep[e.g.][]{hurlburt08}. To simulate the larger scales of 
motion that are likely to occur in the solar convection zone, a large-eddy simulation (LES) model is employed. The scales that are not explicitly
computed in these simulations are parametrized and included in a sub-grid scale model of turbulent transport and diffusion. The equations of 
motion solved in CSS are:

\vspace{-0.3truein}
\begin{center}
   \begin{eqnarray}
     \frac{\partial \rho}{\partial t} + \nabla \cdot \left( \rho \mathbf{u} \right) & = & 0, \nonumber \\
     \rho \left[ \frac{\partial \mathbf{u}}{\partial t} + \left(\mathbf{u} \cdot \nabla \right) \mathbf{u} \right] & = & -\nabla P + 
     \rho g \hat{\mathbf{r}} + \nabla \cdot \left( \mu \overleftrightarrow{\mathcal{D}} \right) \label{momeqn} + 2 \rho \mathbf{u}\times \mathbf{\Omega}
     + \rho \Omega^2 \mathbf{R}, \label{eqnofmot} \\
     \rho T \left[ \frac{\partial S}{\partial t} + \left(\mathbf{u} \cdot \nabla \right) S \right] & = & -\nabla \cdot \mathbf{q} + \rho T \epsilon_0, \nonumber
   \end{eqnarray}
\end{center}

\vspace{-0.1truein}
\noindent where the symbols $\rho$, $\mathbf{u}$, $P$, $T$, $S$, and $\overleftrightarrow{\mathcal{D}}$ are the density, velocity, pressure, 
temperature, specific entropy, and viscous stress tensor respectively; the equation of state is $P = \rho^{\gamma} e^{S/C_V}$; 
$\mathbf{\Omega}$ is the angular velocity of the rotating frame; $\kappa_S$ is a turbulent-eddy entropy diffusion and $\kappa_r$ is the thermal 
diffusion; $g$ is the local acceleration due to gravity; $\epsilon_0$ is a source term that carries the required flux through the boundaries. 
Here $\mu$, $\kappa_S$, $\kappa_r$, $g$, and $\epsilon_0$ are functions of radius only. The diffusive energy flux is 
$\mathbf{q} = -\kappa_S \mathbf{\nabla} S  - \kappa_r \mathbf{\nabla} T - q_{R}\hat{\mathbf{r}}$. The granular-scale cooling is contained in $q_{R}$.

\subsection{Numerical Methods}
The equations of motion (Eqn. \ref{eqnofmot}) are evolved on a uniform spatial mesh. Temporal discretization is accomplished using an explicit 
fourth-order accurate Bulirsch-Stoer time-stepping scheme. The spatial derivatives are computed using a modified sixth-order compact finite 
difference scheme \citep{lele92}. A 3-D domain decomposition divides the full spatial mesh into sub-domains. The boundary information
necessary to compute spatial derivatives is passed between nearest-neighbor sub-domains using MPI, while computations are shared among the master and
slave cores within a supercomputer node using OpenMP.

To begin to understand the dynamics of the near-surface shear layer, we have constructed numerical simulations in spherical shell segments that 
encompass most of the near-surface shear layer and some of the deep interior. A stellar evolution code is used to establish a realistic initial 
stratification for the simulations. Since a perfect gas is assumed, the He and H ionization zones can not currently be simulated. Thus, the upper
boundary ($r_2$) is taken to be $0.995 \, R_{\sun}$ in order to exclude these zones. The lower radial boundary ($r_1$) is the radius 
which corresponds to a given density contrast ($\Delta \rho = \rho_2/\rho_1$, where $\rho_1=\rho(r_1)$ and $\rho_2=\rho(r_2)$). The momentum and 
entropy diffusivities ($\mu$ and $\kappa_s$) are calculated based upon the desired Rayleigh number ($Ra$) at $r_2$ and so that the Prandtl number
($Pr$) is unity throughout the domain. We choose to model these sub-grid-scale diffusivities based upon mixing-length theory, so that the diffusivities 
become functions of radius.

\section{Modeling Granular Convection\label{sect3}}
The strong downflows that occur at edges of a granule drag more dense, lower entropy fluid into the interior. Hence, in our simple model we assume
that the net mass flux is zero and the net entropy flux is negative when integrated over the volume of a granule. It is also assumed that the 
collective action of all the granular-scale dynamics contained within our domain transports the required energy flux through the upper boundary. 
Therefore, we have modeled the granules as cooling events that are advected with the flow, last for an average of $10$ minutes, and have as small 
a spatial scale as the grid and diffusion will allow. The spatial structure of a cooling event is evident in Eqn. \ref{eqn2}. The amplitudes of the
cooling events are chosen at random from an exponential distribution with a mean determined by Eqn. \ref{eqn3}. The horizontal extent of the cooling 
events is normally distributed with a mean angular size of $2 \Delta \phi$ ($\langle \sigma \rangle$) in $\theta$ and $\phi$. The spread of the radial
exponential is fixed to $\sigma_r = 2 \Delta r$ to speed computation. The cooling function is thus

\vspace{-0.5truein}
\begin{center}
  \begin{equation}
    \epsilon = -\nabla \cdot \left(q_R\hat{\mathbf{r}}\right) = \left[\frac{2}{r} + \frac{1}{\sigma_r} \right]
                   e^{-(r_2-r)/\sigma_r} \sum_{n=1}^{N(t)} Q_n(t) e^{-\left[(\theta-\theta_n(t))^2+(\phi-\phi_n(t))^2\right]/2\sigma_n^2}, \label{eqn2}
  \end{equation}
\end{center}

\vspace{-0.1truein}
\noindent where $Q_n(t)$ is the amplitude of the cooling event at time $t$, and the cooling events are also horizontally advected with the flow:
$\theta_n(t)=\theta_n^0+\frac{\Delta t u_{\theta}}{r_2}$, and $\phi_n(t)=\phi_n^0+\frac{\Delta t u_{\phi}}{r_2}$.
 
A mean number of cooling events ($\langle N \rangle$), with ($\langle \sigma \rangle$), determines the mean strength 

\vspace{-0.35truein}
\begin{center}
  \begin{equation}
    \langle Q \rangle = \frac{(\phi_2-\phi_1)\left(\cos{\theta_1}-\cos{\theta_2}\right)}{2 \pi \langle N \rangle \langle \sigma \rangle^2} 
    \left[\frac{L}{4 \pi r^2}+\kappa_S\frac{\partial S}{\partial r}+\kappa_r\frac{\partial T}{\partial r} \right]_{r_2}. \label{eqn3}
  \end{equation}
\end{center}

\vspace{-0.1truein}
\noindent Cooling events have a mean lifetime of $10$ minutes, but can deviate significantly from this as the lifetimes are normally distributed 
as well. When a cooling event has reached the end of its lifetime, it is deleted and at least one new cooling event is generated. A uniform 
distribution is used to choose how many new events are generated. Similarly, at each time step there is also a chance to add new events. In future 
models, improved statistics for the granular scale cooling based upon the surface convection simulations of \citet{tramp07} will be implemented.

\begin{table}[htb!]
   \small
   \begin{center}
   \begin{tabular}{ccccc}
   \multicolumn{5}{c}{Table 1: Properties of the Computational Domain} \\
   \hline
   & \textit{Case 1} & \textit{Case 2} & \textit{Case 3} & \textit{Case 4} \\
   \hline
   $\langle N \rangle$ & $0$ & $250$ & $500$ & $1000$ \\
   $\langle Ra \rangle$ & $5.98 \times 10^6$ & $7.63 \times 10^6$ & $7.90 \times 10^6$ & $7.94 \times 10^6$ \\
   $\langle Re \rangle$ & $133$ & $152$ & $151$ & $150$ \\
   $\langle Ro \rangle$ & $13.3$ & $17.7$ & $16.9$ & $16.6$ \\
   \hline
   \end{tabular}
   \caption{For each case, the depth of the convection zone as modeled is $60$~Mm, $\Delta \rho$ is $400$, the angular size is $20^{\circ}\times20^{\circ}$,
     with a resolution of $128\times256\times256$. $\langle N \rangle$ is the mean number of cooling sites. All cases were run with the same diffusivities
     and initial background. $\langle Ra \rangle$, $\langle Re \rangle$, and $\langle Ro \rangle$, are the average Rayleigh, Reynolds, and Rossby 
     numbers in the domain. \label{tab1}}
   \end{center}
   \vspace{-0.4truein}
\end{table}

\begin{figure}[htb!]
  \begin{center}
    \plotone{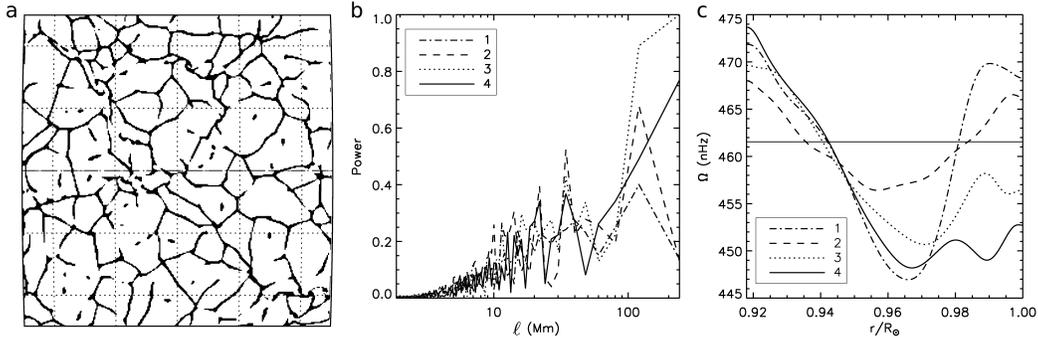}
    \vspace{-0.1truein}
    \caption{(\textbf{a}) A $20^{\circ}\times20^{\circ}$ snapshot of radial velocities ($u_r$) taken at $0.99 \, R_{\sun}$ in case 2: downflows are black, 
      upflows are white. (\textbf{b}) Horizontal scale $\ell$ in $u_r$ power spectra averaged in time and latitude from cases 4 (solid), 3 (dotted), 2 
      (dashed), and 1 (dash-dotted). (\textbf{c}) Horizontally averaged rotation rate $\Omega (r)$, lines as above. \label{fig1}}
    \vspace{-0.3truein}
  \end{center}
\end{figure}

\section{Nature of the Convection\label{sect4}}
In a typical simulation, the convective cells are broad upflows bordered by strong, narrow downflows as seen in Figure \ref{fig1}a. The 
upflows are warm and the downflows are cool relative to the mean temperature. The typical radial velocities in Figure \ref{fig1}a are between
$-1400$~$\mathrm{ms^{-1}}$ for downflows and $400$~$\mathrm{ms^{-1}}$ for upflows. The spatial scales present in the four cases can be seen in power spectra
(Figure \ref{fig1}b). The supergranule-like scales in \textit{Case 1} are about $40$~Mm, just outside the accepted range for supergranules. A spherically 
symmetric cooling is used in \textit{Case 1} which leads to horizontal scales near the surface that are primarily selected by the geometry of the domain. In 
cases with granular-scale cooling, the size scales seen in the simulation decrease due to the granular-scale forcing as evidenced by the leftward 
shift in the power of the peaks below $40$~Mm in Figure \ref{fig1}b. In \textit{Cases 2-4}, supergranular scales become much more prevalent with
power increasing by 50-100\% in the band between $10$~Mm and $35$~Mm. In \textit{Case 2}, a prominent peak occurs around $10$~Mm due to the relative
paucity of cooling sites. Since there are only 250 cooling sites on average, each site necessarily induces a strong downflow, which ensures that the
enthalpy flux is sufficient to carry out the solar flux. Furthermore, there are typically 15 sites at a single longitude or latitude so that the 
average distance between them is $10$-$15$~Mm. Thus, since each cooling site is likely to be a strong downflow, power is present on the $10$~Mm 
scale. Such a peak does not occur to the same extent in \textit{Case 3} or \textit{Case 4} because the probability of inducing a strong downflow at a 
cooling site decreases by factor between 2 to 4. However, a larger $\langle N \rangle$ reduces the average distance between cooling sites and thus allows
for more interaction of the flows generated at the granular scales due to the advection and coalescence of the cooling sites.

As $\langle N \rangle$ is increased, the granular scale cooling disrupts the mean horizontal flows in the upper 20\% of the domain which reduces the 
mean rotation rate as seen in Figure \ref{fig1}c. The reason the mean flows are affected is that when a cooling site forms in an upflow, it begins 
cooling nearby gas causing contracting horizontal flows and a downflow. So, the net effect of this cooling site is to reduce the horizontal flows 
within the upflow as well as slow the upflow's ascent as it has been cooled. Hence, as $\langle N \rangle$ increases the average number of cooling 
sites per upflow increases thus decreasing the mean horizontal flows as in Figure \ref{fig1}c. The radial boundaries are currently impenetrable and 
stress-free leading to overturning of upflows at the top and downflows at the bottom of the domain. This is the primary reason the rotation rate at 
the bottom does not flatten out and why there is an unrealistic shear layer near the surface. Future simulations will have inflow-outflow radial 
boundaries which should eliminate these effects.

\acknowledgements
This work has been supported by Lockheed Martin grant 8100000527 and NASA grants NNG05GI24G and NNX08AI57G. The computations were carried out
on Pleiades at NASA Ames with SMD grant g26133.
\vspace{-0.1truein}

\bibliographystyle{hurlburt}
\bibliography{hurlburt}

\end{document}